\documentclass[11pt,fleqn]{article} %%%ENGLISH

\usepackage{latexsym}
\usepackage{amssymb,amsfonts}
\usepackage[dvips]{graphicx}
\usepackage{euscript} %%%% use with option \mathcal{}
\usepackage{bm}        %% bold mathematics
\usepackage{amsmath}   %% non gradito a latex2html!

\usepackage{amsmath}

\catcode`\@=11
\@addtoreset{equation}{section}
\def\theequation{\arabic{section}.\arabic{equation}}
\def\appendix{\renewcommand{\thesection}{\Alph{section}}\setcounter{section}{0}
              \renewcommand{\theequation}
            {\mbox{\Alph{section}.\arabic{equation}}}\setcounter{equation}{0}}
\def\maketitle{\thispagestyle{empty}\setcounter{page}0\newpage
                \renewcommand{\thefootnote}{\arabic{footnote}}
                  \setcounter{footnote}0}
\renewcommand{\thanks}[1]{\renewcommand{\thefootnote}{\fnsymbol{footnote}}
               \footnote{#1}\renewcommand{\thefootnote}{\arabic{footnote}}}

\renewcommand{\title}[1]{\begin{center}\Large\bf #1\end{center}\rm\par\bigskip}
\renewcommand{\author}[1]{\begin{center}\Large #1\end{center}}
\newcommand{\address}[1]{\begin{center}\large #1\end{center}}

\def\dinfn{\smallskip Dipartimento di Fisica, Universit\`a di Trento\\
                           and Istituto Nazionale di Fisica Nucleare,\\
                                   Gruppo Collegato di Trento, Italia}

\def\Idinfn{\address{\dinfn}}

\newcommand{\email}[1]{e-mail: \sl #1@science.unitn.it\rm}

\newcommand{\femail}[1]{\thanks{\email{#1}}}
\newcommand{\pacs}[1]{\smallskip\noindent{\sl PACS numbers:
                       \hspace{0.3cm}#1}\par\bigskip\rm}
\def\babs{\hrule\par\begin{description}\item{Abstract: }\it}
\def\eabs{\par\end{description}\hrule\par\medskip\rm}
\renewcommand{\date}[1]{\par\bigskip\par\sl\hfill #1\par\medskip\par\rm}

\renewcommand{\vec}[1]{{\bf #1}}       %%%  vectors in bold
\newcommand{\ca}[1]{{\cal #1}}         %%%  calligraphic
\def\hs{\qquad}               %%%  horizontal space
\def\nn{\nonumber}            %%%  no number for eqnarray
\def\beq{\begin{eqnarray}}    %%%  begequation/eqnarray
\def\eeq{\end{eqnarray}}      %%%  endequation/eqnarray
\def\at{\left(}               %%%  open (
\def\aq{\left[}               %%%  open [
\def\ct{\right)}              %%%  close )
\def\cq{\right]}              %%%  close ]
\def\R{{\hbox{{\rm I}\kern-.2em\hbox{\rm R}}}}   %%% real numbers
\def\H{{\hbox{{\rm I}\kern-.2em\hbox{\rm H}}}}   %%% Hilbert space
\def\N{{\hbox{{\rm I}\kern-.2em\hbox{\rm N}}}}   %%% natural numbers
\def\C{{\ \hbox{{\rm I}\kern-.6em\hbox{\bf C}}}} %%% complex numbers
\def\Z{{\hbox{{\rm Z}\kern-.4em\hbox{\rm Z}}}}   %%% integers numbers
\renewcommand{\Im}{\mathop{\rm Im}\nolimits}       %%% Imaginary
\def\al{\alpha}

\def\om{\omega}

\begin{document}
%\tableofcontents       %%%%%%   index of section

\title{Thermodynamical properties of hairy black holes in $n$
  spacetimes dimensions}
\author{Mario Nadalini \femail{nadalini},
Luciano Vanzo\femail{vanzo} and
Sergio Zerbini\femail{zerbini}}
\Idinfn

\babs The issue concerning the existence of exact black hole solutions
in presence of non vanishing cosmological constant and  scalar
fields is reconsidered. With regard to this, in investigating no-hair
theorem violations, exact solutions of gravity  having as a source
an interacting and conformally coupled scalar field are revisited
in arbitrary dimensional non asymptotically flat space-times. New and known
hairy black hole solutions are discussed. The thermodynamical properties
 associated with these solutions are investigated and the
invariance  of the black hole entropy with respect to different conformal
frames is proven.
\eabs

\pacs{04.70.Bw, 04.70.Dy}

\section{Introduction}

It is known that the no-hair conjecture in black hole physics may be
violated by the presence of scalar fields. For example, the existence
of hairy black holes minimally coupled to a scalar has been
considered and studied in
\cite{Henneaux:2004zi,Martinez:2004nb,Henneaux:2002wm,Hertog:2004dr,
Koutsoumbas:2006xj}   
and even for a phantom field \cite{Bronnikov:2006fa}. A four-parameter
family of asymptotically dS, flat and AdS solutions, are reported in
\cite{Bronnikov:2007ws}. Time dependent black holes and their
interpretation within the AdS/CFT duality are reported in
\cite{Bak:2007qw}. Hairy black hole solutions where a phase
transition occur near the horizon have been constructed too
\cite{Gubser:2005ih}, and a study of black holes in Brans-Dicke-Maxwell
theory is in \cite{Cai:1996pj}. 
Then there is the case of conformally coupled non  self-interacting or
self-interacting scalar field. For vanishing cosmological constant and
no interaction, the
first solution was found in \cite{beck}. This solution has some
unphysical features,
like the instability (see for example, \cite{bro} and the recent survey
 \cite{betta0}, but also \cite{McFadden:2004ni}).
However, when a non vanishing cosmological constant is taken into
account, other hairy black hole solutions have been found. Among
them, we would like to mention the black hole solution in $n=3$
dimensions  for a non self-interacting conformally coupled scalar
field in $AdS$ \cite{MZ}, and the $n=4$ solution of the same type
presented in \cite{MTZ}. Further solutions have been presented and
discussed in \cite{Martinez:2004nb,betta}. At the same time,
intereting no hair theorems have recently been proved for $\Lambda>0$,
but without conformal coupling and for convex scalar potentials
\cite{Bhattacharya:2007ap}.  

The main motivation for looking for new exact BH solutions in presence of
scalar fields stems from  the renewed interest in BH solutions with non
vanishing cosmological constant. For instance, the important question
of their stability was recently considered in \cite{Dotti:2007cp}. 
This motivations  was partly triggered by the work by McFadden and
Turok \cite{McFadden:2004ni} concerning the two brane RS-type model,
where the low energy effective
dynamics contain scalar hairy BH solutions, the scalar here being the
radion field which describes dynamically the separation of the two
3-branes. 

We also have  to recall a recent rather unexpected development:
Maeda et al. \cite{Torii:2001p} and Winstanley \cite{betta0})
proved the existence of BH solutions with scalar hairs in
asymptotically non flat space-times AdS and dS. Later Martinez et
al. \cite{martinez} found explicitly some  of these hairy BHs.

In this paper, we will revisit  conformally coupled
self-interacting scalar field in asymptotically $n$ dimensional dS
and AdS space-times.

As already mentioned, the interest being related to the fact that
recently, McFadden and Turok  have solved a low energy brane
effective theory  recasting the model in the form of Einstein
gravity plus a  conformally coupled scalar field. Within this new
framework, they  claim that the old instabilities  are no longer
present.
 The conformal, quartic self-interaction  arises from branes with
cosmological constants.

The presence of  conformal coupling  for the scalar field is
equivalent to  work in the so called Jordan frame, traditionally
defined as that frame in which there is no direct interaction between
the scalar field and the matter fields.
In the Jordan frame, if a BH solution exists, it turns out that
thermodynamic properties of the BH solution is
 quite different  from the standard ones and typically the entropy is
 no longer given by the Area Law.
Furthermore, its computation is non trivial.

However, there exists a general approach to this problem and the
BH entropy may be computed by means of Noether charge method,
introduced by Wald \cite{Wald}. This method is very powerful and
has a geometric origin, but sometimes it does not guarantee the
positivity of the entropy. This is an issue one just encounters
with scalar tensor theories with non minimal coupling, though not only 
with them. A very general and powerful approach combining the Noether
charge method and the quasi-local description of gravity due to Brown
and York has been presented in
\cite{Creighton:1995be,Creighton:1995au}. The former uses an
$n$-dimensional Lagrangian density with matter and dilaton satisfying
the assumptions of scalar-tensor theories we will discuss in the next
Section.         

Recall that for  theory of gravity of general type, described by
a Lagrangian density depending on $f(R)$, $R$ being the Ricci
scalar curvature, the Noether charged method,  gives a simple
expression for the BH entropy (modulo a possible non trivial
constant)

\beq S_{BH}=\frac{V_Fr_H^{n-2}}{4 G}f'(R)(r_H)= \frac{A_H}{4
G}f'(R)(r_H)\,. \eeq Here, $V_F$ is the measure of the fundamental
domain of the horizon manifold and $r_H$ the horizon radius (ex.
for $S_2$, $V_F=4\pi$). As a result, since $f'(R)$ might be negative,
negative BH entropies could  appear.

In principle, in the Jordan frame, the Einstein Eqs. are much more
complicated,
due to the  presence  of  non minimal coupling with the scalar field.
Thus, one may pass to
the so called  Einstein frame, where the scalar field is  minimally
coupled to  gravity and $f(\hat{R})=\hat{R}$. In this related
conformal frame,  one has
related  BH solutions, and the BH entropy is simply given by the
area law \beq S_{BH}=\frac{A_H}{4 G}\,. \label{al} \eeq However,
as soon as the BH thermodynamical quantities are concerned, it is
not difficult to prove (see  Section III) that there is no difference in
dealing with the Jordan  or Einstein frame.

The outline of the paper is as follow. In Sec.~II we outline some
general properties of the scalar tensor
theories we are interested in. In Sec. III we show the
conformal invariance of Hawking temperature and black hole entropy
with respect to conformally related frames. In Sec. IV we revisit
the existence of n-dimensional black hole solutions in
asymptotically dS and AdS space-times. In Sections V, VI, and VII the
solutions are examined in details and  new solutions are presented.
In Sec. VIII, we discuss the
thermodynamics of the BH solutions, with a specific proposal for
dealing with possible negative entropies and we conclude with Sec. IX.

\section{Scalar-tensor theories}

The task of
constructing dynamical models involving conventional gravity and
a scalar field as well, offers in principle unlimited
possibilities (see the textbooks \cite{maeda-fujii} on scalar-tensor
theories (ST); see also \cite{fara} and T.~Clifton's recent dissertation
\cite{clifton} for a very extensive discussions of ST in
cosmology). Their number is considerably reduced if we allow only
for second order field equations to be derivable from an invariant
action principle. So let us start with the following scalar-tensor
theory of gravity, written in the Jordan
frame \cite{Bergmann:1968v,Nordtvedt:1970uv,Wagoner:1970vr}, and
employing reduced Planck units such that $8\pi G=1$,
\begin{eqnarray}
\label{a}
I=\int d^{n}x \, {\sqrt {-g}} \left[
\frac {1}{2}F(\phi)R -\Lambda
-\frac {1}{2} \left( \nabla \phi \right) ^{2}
 -V(\phi ) \right] , \label{jf}
\end{eqnarray}
where $F(\phi)$ is a positive function,  $R$ is the
Ricci scalar curvature and $\Lambda$ is the cosmological
constant. Below we shall consider the special case
$F(\phi)=1-\xi\phi^2$, where $\xi$ is a positive coupling parameter.
The scalar amplitude will be measured in Planck units, but we recall
here that $G=M_{pl}^{2-n}$ in $n$ space-time dimensions. In
scalar-tensor gravity there really is not a ``constant Newton
 constant'', since the effective $G$ is a field; from Eq.~\eqref{jf}
 we see that
this effective gravitational coupling is $G=1/8\pi F$, although
this is not the one which is measured in a Cavendish-like
experiment, since the scalar field will contribute to the
inter-particle force. Coupling constants for scalar-scalar and
scalar-gravity interactions are contained in the functions
$F(\phi)$ and $V(\phi)$, together with any possible mass term for
the scalar field. In principle a function $Z(\phi)$ could have
been introduced in front of the kinetic term, but it can be set to
$\pm1$ by a field redefinition without changing the conformal
class of the metric. If this  is also changed by some conformal
transformation, then one can go to a so called Einstein frame
where the action takes the usual form of general relativity and
the matter action depends on $\phi$, but we prefer to stick to the
Jordan frame here from. So if a matter action is added to $I$, say
$I_M[\psi_M,g_{ab}]$, it is assumed (with Dicke) that is not
directly coupled to the scalar field. This is consistent with the
equivalence principle and the fact that the ratio $\al$ between
the couplings to matter of scalar and tensor field is bounded by
about $10^{-3}$ by solar-system experiments. In any case, the only
coupling of $\phi$ to matter that can be removed from the action
to conform with Dicke's assumption, is clearly of the form
$I_M[\psi_M,\om^2(\phi)g_{ab}]$, which means that $\phi$ will have
matter acting as a source only through the trace of its
energy-momentum tensor\footnote{In particular, a dilaton coupling
to a
  $U(1)$ gauge filed like $h(\phi)F^2$ is forbidden under Dicke
  assumption.}.

The field equations following from \eqref{a} take the well known form
\beq\label{fe1}
F(\phi)\at R_{\mu\nu}-\frac{1}{2}Rg_{\mu\nu}\ct &+& g_{\mu\nu}\nabla^2
F(\phi)-\nabla_{\mu}\nabla_{\nu}F(\phi)
\nn \\
&-&\nabla_{\mu}\phi\nabla_{\nu}\phi+\frac{1}{2}(\nabla\phi)^2g_{\mu\nu}+
(\Lambda+V(\phi))g_{\mu\nu}=0
\eeq
\beq\label{fe2}
\nabla^2\phi+\frac{1}{2}F^{'}(\phi)R-V^{'}(\phi)=0
\eeq
If matter is present its energy tensor should be added to the right
hand side of \eqref{fe1}, but Eq.~\eqref{fe2} will remain
unaffected. If the scalar is a constant, say $\phi=\phi_0$, the metric
solves the ordinary Einstein's equations with an effective
cosmological constant
$\Lambda_{eff}=(\Lambda+V(\phi_0))/F(\phi_0)$. The constant $\phi_0$
can then be related to the Newton constant as measured by a Cavendish
experiment as follow. If the field is slowly varying we can expand
the action \eqref{a} around $\phi_0$ and keeps first order terms; we
obtain the action
\[
I=\frac{1}{2}\int\at
 (F_0+F^{'}_0\sigma)R-(\nabla\sigma)^2-V(\phi_0+\sigma)-\Lambda\ct|g|^{1/2}
d^4x
\]
We work in $4D$ for simplicity and define $F_0=F(\phi_0)$,
$V_0=V(\phi_0)$, etc. In a
Cavendish experiment we are on space-time scales much less than
 $\Lambda_{eff}^{-1/2}$, the effective vacuum energy of the
theory. Redefining $\varphi=F_0+F^{'}_0\sigma$ we obtain the
linearized action of Brans-Dicke theory with BD parameter
$\om=F_0/(F^{'}_0)^2$ and a scalar potential. The gravitational
coupling of this theory in the limit of a massless scalar is (see,
e.g. \cite{weinbook})
\[
G=\frac{1}{8\pi\varphi_0}\frac{2\om+4}{2\om+3}
\]
so we obtain $G$ for the massless (or nearly massless) scalar-tensor
theory
\beq\label{G}
G=\frac{1}{8\pi F_0}\frac{2F_0+4(F^{'}_0)^2}{2F_0+3(F^{'}_0)^2}
\eeq
In the opposite limit where the kinetic term is unimportant while the
potential term dominates, $G$ reduces to the first factor in
\eqref{G}, but it still depend on $\phi_0$. An example is the theory
with Lagrangian
density $\ca L=\phi R+V(\phi)$, which is just modified gravity
with $\ca L=f(R)=V(\phi(R))-\phi(R) V^{'}(\phi(R))$, where
$\phi(R)$ is the solution of the equation $R=-V^{'}(\phi)$.
Thus $\phi_0$, the constant solution, can be measured by a Cavendish
experiment in a slowly varying field. Since $\phi_0$ is related to
other parameters of the model by the field equations, \eqref{G} is an
important constraint on these parameters. In particular, it implies
the obvious requirement $F_0>0$. For the theory considered in this
paper, corresponding to $F(\phi)=1-\xi\phi^2$, one has a possible
pathology when the  amplitude $\phi\sim\xi^{-1/2}$ is of order of the
Planck masses, assuming a positive $\xi$ of order one.

\section{Equivalence of Hawking temperature and entropy in conformally
  related frames}

It is known that the surface gravity of a stationary black hole is
invariant under conformal transformations of the metric that are
the identity at the infinity \cite{Jacobson:1993pf}. An alternative and
simpler proof for static black holes, with an additional result
concerning the entropy, goes as follows. Assume that exists a static
black hole solution 
\beq
ds^2=-A(r)dt^2+\frac{dr^2}{B(r)}+r^2d\Sigma^2\,. \nonumber
\eeq

Here $\Sigma$ is a  maximally symmetric space with volume $V_F$,
 representing the event horizon manifold located at $r_H$, where
 $A(r_H)=B(r_H)=0$,
with  $A'(r_H)\neq 0$ and $B'(r_H)\neq 0$. The Hawking temperature
can be computed by means of (see, for example
 \cite{Parikh:1999mf,Angheben:2005rm,svmp,Medved:2005yf,Kerner:2006vu,mitra}
and references cited therein  for a dynamical derivation) \beq
T_H=\frac{\sqrt{A'(r_H)B'(r_H)}}{4\pi}\,, \eeq while the entropy,
computed by means of the Noether charge method devised by Wald,
 reads\footnote{Recall that $4G=1/2\pi$ with our units.}
 \beq S_{BH}=2\pi V_F r_H^{n-2}F\left(\phi_H\right)\,. \label{s} \eeq
where $\phi_H=\phi\left(r_H\right)$. This formula can also be obtained
 by quasi-local methods \cite{Creighton:1995au}, but here too the
 positivity of $S_{BH}$ cannot be guaranteed without additional
 assumptions. We note that the condition for the
 positivity of the Newton constant, which were $F(\phi)>0$ for the relevant
 range of $\phi$, is equivalent to that requiring the positivity of the
 entropy, and also that $S_{BH}$ is controlled by the effective
 gravitational coupling $G_e=1/8\pi F(\phi_H)$ computed on the
 horizon, not by the coupling \eqref{G}. This fact is similar to the classical
 analogue of the non-renormalization theorem of black hole entropy,
 namely the fact that the effects of high energy degrees of freedom on
 the entropy of black holes are just the same ultraviolet effects that
 renormalize the Newton constant.

Now pass from the Jordan frame to the
 Einstein frame by the conformal transformation
\beq
\hat{g}_{\mu \nu}=\Omega^{\frac{2}{n-2}} g_{\mu \nu}\,,
\eeq
with
\beq
\Omega=F(\phi(r))\,.
\eeq
The BH solution in the Einstein frame  becomes
\beq
d\hat{s}^2=-dt^2 \hat{A}(r)+\frac{dr^2}{\hat{B}(r)}+\hat{r}^2d\Sigma^2\,,\nn
\eeq
\beq
\hat{A}(r)=F(\phi)^{\frac{2}{n-2}}A(r)\,, \nonumber
\eeq
\beq
\hat{B}(r)=F\phi)^{-\frac{2}{n-2}}B(r)\,, \nonumber
\eeq
\beq
\hat{r}=F(\phi)^{\frac{1}{n-2}}r\,.
\label{r}
\eeq
Here (Einstein frame) the  scalar field is minimally coupled  with a 
complicated potential. If $\Omega$ and its first derivatives are well
behaved on the horizon, the coordinate location of the horizon will be
unaffected by the conformal transformation, and the Einstein frame
entropy reads
\beq
\hat{S}_{BH}=2\pi V_F (\hat{r}_H)^{n-2}\,. \nonumber
\eeq
As a result, from (\ref{r}),
\beq
\hat{S}_{BH}=S_{BH}\,. \nonumber
\eeq
The Hawking temperature in the  Einstein frame is
\beq
\hat{T}_H=\frac{\sqrt{\hat{A}'(r_H)\hat{B}'(r_H)}}{4\pi}\,. \nonumber
\eeq
Since
\begin{gather}
\hat{A}'(r_H)=A'(r)F(\phi)^{\frac{2}{n-2}}+O(r-r_H)\,, \nonumber\\
\hat{B}'(r_H)=B'(r)F(\phi)^{-\frac{2}{n-2}}+O(r-r_H)\,, \nonumber
\end{gather}
one immediately obtains
\beq
\hat{T}_H=T_H\,. \nonumber
\eeq

We have proved this result: If BH solutions exist, the Hawking
temperature and BH entropy are invariant quantities under conformal
transformations with factors that have finite values together with
finite first order derivatives near and on the event horizon. In
particular, they are invariant with respect to different conformally
related frames, an important point spelled out from different
perspectives also in  \cite{Cai:1996pj,Chan:1996sx}.

\section{Static and spherically symmetric solutions}

In this Section, we will revisit the existence of static and
spherically symmetric solutions which are asymptotically de Sitter
(dS) and Anti-de Sitter (AdS) working in the Jordan frame, for the
choice $F(\phi)=1-\xi\phi^2$. Since  a generally
valid gravitational constant in ST does not exist, we make use of
conventional Planck units $8\pi G_0=1$, where $G_0^{-1/2}$ is the
Planck mass as measured in a vacuum with a vanishingly small scalar
field. Below we relate it to the Planck mass in a vacuum with a
constant non zero value of $\phi$.

\subsection{Vacuum solution}
To begin with, recall the action for a scalar field $\phi$ with
non-minimal coupling and self-interaction $V(\phi)$, in $n-$dimensions
in the Jordan frame: 
\begin{eqnarray}
\label{aa}
I=\int d^{n}x \, {\sqrt {-g}} \left[
\frac {1}{2}\left( R -2\Lambda \right)
-\frac {1}{2} \left( \nabla \phi \right) ^{2}
-\frac {1}{2} \xi R \phi ^{2} -V(\phi ) \right]\,.
\end{eqnarray}
The special value, $\xi_c=(n-2)/4(n-1)$ and
$V(\phi)\sim\phi^{2n/(n-2)}$, gives a scalar field theory 
which is conformally invariant on a curved background  with a non
dynamical metric. The Eqs. of motions read (see  for example \cite{betta})
or  \eqref{fe1}, \eqref{fe2}
\begin{eqnarray}
\label{e1}
 \left( 1- \xi \phi ^{2} \right) G_{\mu \nu }
+g_{\mu \nu } \Lambda
&=&
\left(1 - 2\xi \right) \nabla _{\mu } \phi \nabla _{\nu }
\phi
+\left(2 \xi -\frac {1}{2} \right) g_{\mu \nu } \left(
\nabla \phi \right)^{2}
\nonumber
\\
&&-2\xi \phi \nabla _{\mu } \nabla _{\nu } \phi
+2\xi g_{\mu \nu } \phi \nabla ^{2}  \phi
- g_{\mu \nu } V(\phi ),
\\
\label{e2}
\nabla^2 \phi - \xi  R\phi -
\frac {dV}{d \phi } &=&0.
\end{eqnarray}
where $G_{\mu \nu }=R_{\mu \nu }-\frac{R}{2}g_{\mu \nu }$ is the
Einstein tensor. Taking these Eqs. of motion into account and
taking the trace, one gets
\begin{eqnarray}\label{e3}
R=-\frac{ 2(n-1)(\xi-\xi_c)(\nabla \phi)^2
+2\xi(n-1)\phi\frac{dV}{d\phi}-n(V(\phi)+\Lambda) }
{n/2-1+2(n-1)\xi(\xi-\xi_c) \phi^2 }.
\end{eqnarray}
The scalar curvature, and thus the effective potential in
Eq.~\eqref{e2}, has a pole at the critical
field \cite{Hosotani:1985at}
\[
\phi_{c}^2=\frac{\xi_c}{\xi(\xi_c-\xi)}
\]
which is only defined for $0<\xi<\xi_c$; its $4D$ cousin is
significant for the stability analysis of the constant
solutions \cite{Hosotani:1985at}. The other critical points are at
$\widehat{\phi}=\pm|\xi|^{-1/2}$, where the effective Newton constant
vanishes, and are important in the phase space
structure of the theory \cite{Amendola:1990nn,Barroso:1991aj}.
Substitution of \eqref{e3} into Eq.~\eqref{e2} yields
\beq\label{e4}
\nabla^2\phi-U^{'}(\phi)+\frac{\phi}{\phi^2-\phi_{c}^2}(\nabla\phi)^2=0
\eeq
\beq\label{e5}
U^{'}(\phi)=-\frac{\phi_c^2}{\phi^2-\phi_c^2}\aq(1-\xi\phi^2)V^{'}(\phi)+
\frac{2n\xi}{n-2}\phi(V(\phi)+\Lambda)\cq
\eeq
To recover general relativity we require that a constant $\phi$, say
$\phi_0$, be a solution of the field equations.
With this ansatz, the Eqs. of motion reduce to \beq
\at 1-\xi \phi_0^2 \ct\at \frac{n-2}{2} \ct R_0
=\Lambda+V(\phi_0)\,, \eeq
\beq \xi
R_0\phi_0+\frac{dV}{d\phi}(\phi_0)=0
\eeq
where $R_0$ is the constant scalar curvature. These equations
determine $R_0$ and $\phi_0$ of the background manifold, and may or
may not have a solution. If there is one,  the metric will
describe an Einstein manifold, for which one has
\beq
G_{\mu \nu }^{(0)}= \frac{(2-n)}{2n} R_0 g_{\mu \nu }^{(0)}\,,
\label{0} \eeq
but the space will not, in general, be maximally symmetric. Under this
stronger requirement it is clear that the theory will admit Minkowski,
de Sitter and anti-de Sitter vacuums, if only one chooses
appropriately the relevant parameters like $\Lambda$, $V(\phi_0)$,
$\xi$ and so on. The stability of these vacuums
were analyzed by Hosotani \cite{Hosotani:1985at} in four dimensions for
a quartic polynomial potential. A cubic interaction term induces
instability unless $\xi=0$. In higher dimensions, for a potential
\[
V(\phi)=\alpha_n\phi^{2n/(n-2)}+\frac{1}{2} m^2\phi^2
\]
Hosotani's stability criterium  against spatially homogeneous
perturbations gives: for $\xi\leq 0$ or
$\xi\geq\xi_c$  the condition is $2\al_4+m^2\xi>0$ for $n=4$, $\al_3>0$ for
$n=3$ and $m^2\xi>0$ for $n>4$, while for $0<\xi<\xi_c$ and $m^2=0$ the
condition is
\[
\al_n\left(\frac{\xi_c}{\xi_c-\xi}\right)^{2/(n-2)}+\Lambda\xi^{n/(n-2)}>0
\]
Thus a negative $\al_n$ can make sense in curved space.
Next we consider just two examples, a potential with mass term and a
conformally invariant case. Thus we have, to start with,
\[
V(\phi)=\frac{1}{2}\mu^2\phi^2
\]
and the vacuum equations give either $\phi_0=0$ and
$R_0=2\Lambda/(n-2)$, or $R_0=-\mu^2/\xi$ and
$\phi_0^2=(n-2)\mu^2+2\xi\Lambda)/(n-3)\mu^2\xi$. For $n=3$, $\phi_0$
is either zero or it is undetermined.

Let us consider now a
symmetric vacuum solutions with conformal symmetry in the scalar
sector, this means that \beq \xi=\xi_c=\frac{n-2}{4(n-1)}\,, \hs
V(\phi)=\alpha_n \phi^{\frac{2n}{n-2}} \,, \hs
R_0=\frac{2n}{n-2}\Lambda\,. \eeq As a result, we get \beq \xi_c
R_0\phi_0^2=-\frac{2n}{n-2}\alpha_n \phi_0^{\frac{2n}{n-2}}\,,
\eeq which satisfies also the equation of motion for the scalar
field. Thus, we have found the following relation between the
vacuum solutions \beq \xi_c \Lambda=-\alpha_n
\phi_0^{\frac{4}{n-2}}\,. \eeq For example, when $n=3$, \beq
\phi_0^4=-\frac{\Lambda}{8\alpha_3}\,. \eeq If $n=4$ \beq
\phi_0^2=-\frac{\Lambda}{6\alpha_4}\,. \eeq For these values of
$n$, in order to have a positive self-interacting coupling
constant, one has to deal with negative cosmological constant, in
agreement with \cite{vanzo}, but a negative $\al_n$ can still make
sense (this is due to the back reaction of gravity on the scalar
field). A positive cosmological constant is consistent only for
$\alpha_n <0$ \cite{Martinez:2004nb}. For $n >4$, the relation between
$\Lambda$ 
and $\alpha$ involve non manifestly positive quantities.

\subsection{Analysis of the exact solutions}
It is known that if one allows an Einstein space as horizon manifold, a
reasonable ansatz for the BH metric reads (see, for example, \cite{vanzo1})

{\begin{eqnarray}
ds^{2} =- A(r) dt^{2}+\frac{dr^{2}}{A(r)}+ r^{2} d \Sigma^2\,, \nonumber
\end{eqnarray}
where $d\Sigma$ is the metric of (n-2)-dimensional maximally
symmetric Einstein space, whose normalized constant sectional
curvature is $k=0,\pm 1$. Assume also that $\phi(x)=\phi(r)$, and
the conformal invariance of the scalar sector, namely
$\xi=\xi_c=\frac{n-2}{4(n-1)}$ and \beq V(\phi)=\alpha_n
\phi^{\frac{2n}{n-2}}\,. \nonumber \eeq Thus,  Eqs. of motion
reduce to ( $'$ is the derivative with respect to $r$) {\beq\label{m1}
-(1-2\xi_c)\phi'^2+2\xi_c\phi\phi''=0\,,  \eeq \beq\label{m2}
 \phi' A'+\left(\phi''+\frac{n-2}{r} \phi'\right)A-\xi_c R \phi
-\frac{dV(\phi)}{d \phi}=0\,,
%\nonumber
\eeq
\begin{multline}\label{m3}
\frac{n-2}{2r}(1-\xi_c \phi^2)\left[A' -\frac{n-3}{r}(k-A)\right] -
\left(2\xi_c-\frac{1}{2}\right)A\phi'^2+\xi \phi \phi'A'+  \\
-2\xi_c \phi\nabla^2 \phi +V(\phi)+\Lambda=0\,.
\end{multline}
The  first equation is a combination of two independent Einstein
Eqs., one of which reduces to the third equation, while the second
equation is the Eq. of motion for the scalar field.

Within our specific assumptions (the conformal invariance of the
scalar sector), the first Eq., is a  simple and  non linear
equation involving only $\phi(r)$, and the general solution is
\beq \phi(r)=c \at r+r_0 \ct^{1-n/2}\,.\label{p} \eeq with  $r_0$
and $c$  constants of integration.

Once this solution is  plugged into the other two equations, the
second and third ones become  first order differential equations
with non constant coefficients for the unknown left function
$A(r)$: a problem of compatibility arises and the solutions for
$A$  may exist or may not exist. Thus a simple procedure is at
disposal:   solve the second (first order diff.) equation,  then
verify if the obtained  solution satisfies  identically the third
equation. The solutions, when they exist, lead to relations
between the constants of integration $c$ and $r_0$ and the
parameters $\Lambda$ and $\alpha_n$ of the model.

The general solution of the second diff. equation is standard and
 reads \beq A(r)= (r+r_0)^{n/2}r^{2-n}\at k_0+\int
(r+r_0)^{-n/2}r^{n-2}B_n(r) dr \ct\,, \label{2} \eeq where $k_0$
is a further integration constant and \beq
B_n(r)=-\frac{n\Lambda}{(n-2)(n-1)}(r+r_0)-
\frac{4\alpha_n}{(n-2)^2}c^{4/(n-2)}(r+r_0)^{-1}\,. \nonumber \eeq
The integral involving the function $B_n(r)$ may be evaluated for
 generic $\Lambda$ and $n$, but for the sake of simplicity, we prefer to deal
with a definite sign of $\Lambda$ and specific values of $n$.

\section{The $\Lambda < 0$ case}

Since $\Lambda < 0$, we may introduce the length $l$ by means of
$\Lambda=-\frac{(n-2)(n-1)}{2l^2}$.

\subsection{The $n=3$ solution}

For $n=3$, $\xi_c=1/8$ and the potential reads $V=\alpha_3 \phi^6$.
The solution for the scalar field reduces to \beq \phi(r)=\at 8
\frac{r_0}{r+r_0}\ct^{1/2}\,. \eeq A direct calculation gives \beq
A(r)=k_0\frac{(r+r_0)^{3/2}}{r}+\frac{r^2}{l^2}-
r_0^2(3+2\frac{r_0}{r})\Delta\,, \label{3} \eeq where \beq
\Delta=\frac{8\alpha_3c^4}{r_0^2}-\frac{1}{l^2} \eeq If we plug this
solution into the third equation \eqref{m3}, one finds that it is
satisfied when $k_0=0$ and $\Delta=0$ or under the different set of
conditions $k_0=0$ and $c^2=8r_0$. In the first case, \beq
\frac{8\alpha_3c^4}{r_0^2}=\frac{1}{l^2}\,, \hs A(r)=\frac{r^2}{l^2}
\,.\label{6} \eeq We may anticipate that this form of the lapse
function exists for arbitrary $n$.

In the second case, $r_0=\frac{c^2}{8} >0$ and \beq A(r)=
\frac{r^2}{l^2}+\frac{r_0^2}{l^2}(\omega-1)\at 3+\frac{2r_0}{r}\ct
\,, \label{8} \eeq where \beq \omega=512 l^2 \alpha_3 > 0\,.
\nonumber \eeq Recall that one has black hole solution as soon as
$A(r)$ has a  positive root $r_H$ with $A'(r_H) \neq 0$.
The roots are solutions of third order algebraic equation \beq
r^3+3r_0^2(\omega-1) r+2r_0^3(\omega-1)=0 \,. \eeq The related
discriminant  reads \beq D=a^6(\omega-1)^2\omega \,. \eeq One has
two cases. The  first is $\omega=0$, $D=0$, corresponding to zero
scalar self-interaction, the roots are real and the positive
one reads \beq r_H=2r_0\, . \eeq The related solution is \beq
A(r)=\frac{r^2}{l^2}-3\frac{r^2_0}{l^2}-2\frac{r^3_0}{l^2 r} \eeq
In the other case, $ 1>  \omega >0$, $D >0$ and there exists only one
real positive root \beq r_H=r_0f(\omega)=r_0 \at 1-\omega \ct^{1/3}\aq
\at 1+\sqrt\omega \ct^{1/3}+ \at 1-\sqrt \omega \ct^{1/3}\cq\
\nonumber \eeq

\subsection{The $n=4$ solutions}

For $n=4$, the situation changes. First, in order to satisfy the
second differential equation, one finds that \beq
r^2_0=2\alpha_4c^2 l^2\,. \eeq Furthermore, with  choice $k_0=k$,
the third equation gives \beq 0=\frac{6k}{r^4}(c^2-6r_0^2) \eeq As
a result, one has the set of solutions with $c=\pm \sqrt{6}r_0$,
with $r_0 >0$  and  $r_0 <0$. In general \beq A(r)=k
\frac{(r+r_0)^{2}}{r^2}+\frac{r^2}{l^2}\,, \eeq \beq
\frac{\phi(r)^2}{6}=\frac{r_0^2}{(r+r_0)^2}\,. \eeq The
self-interacting coupling constant is  fixed to be \beq
\alpha_4=\frac{l^2}{12}\,. \eeq The solution for $\phi(r)$ is
regular everywhere for $ r_0 <0$.

Furthermore, for $k=1$ (spherical transverse manifold),
$A(r)$ is   a regular  non
vanishing lapse function and the solution is  asymptotically AdS.

 For $k=0$ (locally flat transverse manifold), the metric is
conformally and locally related to Minkowski space-time.

For $k=-1$ (hyperbolic transverse manifold), one has  a BH solution,
since $A(r)$  has real roots, the biggest one being the radius of the
event horizon
\beq
r_H=\frac{l}{2}\at 1+\sqrt{1+4r_0/l} \ct\,.
\eeq

This kind of solutions are conformally related to the solutions
found by Martinez et al \cite{Martinez:2004nb} working in the Einstein frame.

\subsection{A charged $n=4$ black hole solution}

Let us consider the $n=4$ and let us investigate the existence of
a charged black hole solution in the presence of a neutral
conformally coupled scalar field. For $\Lambda > 0$, a
corresponding solution  has recently been found in 
\cite{Martinez:2004nb}. One has to add the Maxwell term to the scalar
gravity 
action and observes that the only non vanishing component of the
EM tensor field is \beq F_{tr}=\frac{Q}{r^2}\,. \eeq The Eqs. of
motion are supplemented by the EM stress tensor, given by
$T_{\mu\nu}^{EM}=\frac{Q^2}{8\pi r^4}\varepsilon g_{\mu \nu}$ with
$\varepsilon=-1$ for $\mu=t,r$ and $\varepsilon=1$ for the other
spatial coordinates. It is easy to see that the first equation of
motion is left unchanged, namely \beq 2\phi'^2-\phi\phi''=0\,,
\eeq and the solution is \beq \phi(r)=\frac{c}{r+r_0 }\,.
\label{10} \eeq As a consequence, with \beq
\frac{r_0^2}{l^2}=2\alpha_4c^2 \eeq the solution of the second
equation is still the same, namely \beq A(r)=k
\frac{(r+r_0)^{2}}{r^2}+\frac{r^2}{l^2}\,. \label{11} \eeq The
third equation, gives the consistency condition regarding the
constants of integration. The result is \beq
6k(c^2-6r_0^2)+\frac{Q^2}{8\pi}=0\,, \eeq As a result, within the
ansatz we are dealing with, $k=0$ solution is not possible, unless
$Q=0$. A solution of this kind, but in the de Sitter space, has
been reported in \cite{Martinez:2004nb}. Very recently
the case $k=-1$, which is a black hole solution, has been reported
in \cite{martinez}. The other solution, namely the one with $k=1$
is a regular asymptotically AdS solution.

\subsection{The solutions for $ n > 4$}

The previous procedure can be applied to the cases  $n=5, 6,...$.
The computation becomes more and more involved, and the results
are the following:

For arbitrary  $n >4$, and with $k \neq 0$, it seems unlikely that
other solutions exist under our assumptions about the form of
the metric (we have not found any up to $n=11$, but were unable to
prove this in general). If $k=0$ and if
\beq
\alpha_n=\frac{(n-2)^2 r_0^2}{8 l^2 c^{\frac{4}{n-2}}}
\eeq}
holds,  then the   unique solution is
\beq
A(r)=\frac{r^2}{l^2}\,.
\eeq
In this case, the self-interacting coupling constant  $\alpha_n$ is
positive
and with the coordinate change $r=\frac{l^2}{y}$, the lapse function
becomes
\beq
A(y)=\frac{l^2}{y^2}\at -dt^2+dy^2+l^2dT^2_{n-2}\ct\,,
\eeq
namely  it represents locally the AdS space-time.

\section{The $\Lambda >0$ case}

If $\Lambda > 0$, we set  $\Lambda=\frac{(n-2)(n-1)}{2l^2}$. The
procedure is formally similar to the previous one. There exist non
trivial black hole solutions for $n=3$ and $n=4$. They  may be
obtained by the solutions found for $\Lambda <0$,  letting $l^2
\rightarrow -l^2$. Besides  event horizons,  also cosmological and
Cauchy horizons appear and negative BH entropies have been claimed to 
exist~\cite{betta}.

For example, for $n=3$, one has $r_0=\frac{c^2}{8} >0$ again and
\beq A(r)= -\frac{r^2}{l^2}+\frac{r_0^2}{l^2}(\omega+1)\at
3+\frac{2r_0}{r}\ct \,, \eeq where \beq \omega=512 l^2 \alpha_3\,,
\eeq the event, cosmological and Cauchy horizons are defined by
the real roots of \beq r^3-3r_0^2(\omega+1)
r-2r_0^3(\omega+1)=0\,. \eeq First if  $\omega=0$, $D=0$
corresponding to zero scalar self-interaction, the roots are  real
and the positive one reads \beq r_H=2r_0\,. \eeq The related
solution is \beq
A(r)=-\frac{r^2}{l^2}+3\frac{r^2_0}{l^2}+2\frac{r^3_0}{l^2 r}\,.
\eeq If $ \omega <0$, $D >0$ and there exists only one real root.
It is positive and reads \beq r_H=r_0 f(\omega)\,, \eeq \beq
f(\omega)=   \at 1+\omega \ct^{1/3}\aq \at 1+\sqrt{-
\omega}\ct^{1/3}+ \at 1-\sqrt{- \omega} \ct^{1/3}\cq\,. \eeq

\subsection{The $n=4$ solution}

In order to have a black hole, one must have  $k_0=1$ and we have
\beq A(r)= \frac{(r+r_0)^{2}}{r^2}-\frac{r^2}{l^2}\,, \eeq \beq
\phi(r)=\frac{\sqrt 6r_0}{(r+r_0)}\,. \eeq
This reduces to a vacuum de Sitter space when $r_0=0$, so we can think
of it as a hairy excitation of de Sitter space. Nevertheless, as we
will see, the
entropy of the cosmological horizon is less than that of pure de
Sitter space. The self-interacting
coupling constant is  fixed to be  negative \beq
\alpha_4=-\frac{l^2}{12} \nonumber \eeq In this case, if $r_0 <0
$, there exist event,  cosmological and Cauchy horizons and the
transverse manifold is a 2-sphere \cite{Martinez:2004nb}. It  has
been proven that this solution is  unstable. If $r_0 >0$, only  a
cosmological horizon exists.

\subsection{A new n-dimensional de Sitter hairy solution}

As last case, it is easy to see that for generic $n$,  and for
$k=0$, if \beq \alpha_n=-\frac{(n-2)^2 r_0^2}{8 l^2
c^{\frac{4}{n-2}}}\, \eeq is satisfied, then there  exists a
solution given by \beq A(r)=-\frac{r^2}{l^2}\,. \eeq We  may
interpret this solution observing that now $r$ is a time
coordinate and $t$ a space coordinate. Introducing  new
coordinates $(T,Y)$ \beq t=\frac{lY}{T_0}\,\hs
r=T_0e^{\frac{T}{l}}\,, \eeq one has \beq ds^2=-dT^2+
e^{\frac{2T}{l}}\at dY^2 + d\vec{ X}^2 \ct \,, \eeq where the
transverse metric $d\vec{ X}^2$ represents the non compact
manifold $R^{n-2}$, then
 $dY^2 + d\vec{ X}^2$ has the metric of $R^{n-1}$.
Introducing spherical coordinates $(R, \theta_1, \theta_2
\ldots)$, the metric can be rewritten as \beq ds^2=-dT^2+
e^{\frac{2T}{l}}\at dR^2 + R^2 dS^2_{n-2} \ct \,. \eeq This is a
de Sitter-like solution in the cosmological synchronous gauge. In
three dimensions it emerges as the formal zero temperature limit of
the Kerr-de Sitter metric \cite{KV} and represents a cilindrical
universe expanding\footnote{Changing $T\to-T$ one gets a contracting
  one.} from a wirelike singularity. The solution is timelike and null
geodesically incomplete in the past. It can also be shown that the
$n=3$ solution is the asymptotic limit
(in time) of the general solution of $3D$ de Sitter gravity with flat
or toroidal spatial topology, a kind of attractor mechanism.
The fact that the zero
temperature state of three-dimensional de Sitter gravity is not a true
ground state makes it hard to give sense to the statistical partition
function of the finite temperature states. In fact it seems that
without further prescriptions we could get negative entropy states.

As anticipated, the solution for the scalar field also becomes
``time-dependent''
\beq
\phi(T)=\frac{1}{\at T_0e^{T/l}+r_0 \ct^{n/2-1}}\,. \eeq Making
use of the standard mapping, \beq T=\hat{t}+\frac{l}{2}\ln \at
1-\frac{\rho^2}{l^2}\ct\,, \eeq \beq
R=\frac{\rho}{(1-\frac{\rho^2}{l^2})^{1/2}}e^{-\hat{t}/l}\,, \eeq
we may pass to the  static de Sitter gauge \beq ds^2=-d
\hat{t}^2(1-\frac{\rho^2}{l^2})+\frac{d\rho^2}{(1-\frac{\rho^2}{l^2})}
+\rho^2 dS^2_{n-2}\,. \eeq In static gauge, the solution for the
scalar field solution is time dependent and assumes the form \beq
\phi(\hat{t}, \rho)= \frac{1}{\at r_0+T_0
\sqrt{1-\frac{\rho^2}{l^2}} e^{\frac{\hat{t}}{l}}\ct^{n/2-1}} \,.
\eeq

This solution may interpreted as a n-dimensional hairy scalar de
Sitter solution.

\section{The non-warped class of black hole solutions}

Broadening, in some sense, our initial ansatz in order to find new
solutions, we may pass to a setting that will allow us to find
metrics that could be regarded as near-horizon approximations of
other more standard solutions. In particular, in the case
explained below we are also able to find a five dimensional
solution, that could serve as
a guideline for finding other solutions in $n\geq5$ dimensions.\\
The ansatz considered here is reminiscent of the solutions
describing extremal limits of black holes (cfr. for example
\cite{Caldarelli:2000wc}):
\begin{equation}
dS^2=-A(r)dt^2+\frac{dr^2}{A(r)}+B^2d\Sigma^2,
\end{equation}
where $B$ is a suitable real constant, possibly depending on the
physical parameters of the solution. We intend for ``non-warped
solution'' the fact that this metric represent a direct product of
two manifolds, the one described by coordinates $(r,t)$ and
$\Sigma$, not a warped one. In the examples that follow, the
unspecified transverse manifold $\Sigma$ will always be the $n-2$
dimensional sphere.

All the following solution support a constant contravariant electric
field $E$. The value of the electric charge is taken to be $e$.

\subsection{The n=3 solution}

In this setting, the solution for the metric and the field turns
out to be
\begin{align}
    A(r)&=\frac{2}{l^2}{(r-r_0)}^2+8\alpha c^4;\\
    \phi(r)&=\frac{c}{{(r-r_0)}^{1/2}}.
\end{align}
The physical constant are related trough $\Lambda=-1/l^2=-e^2$, $B$
remains a free parameter.\\ The constants $c$ can always be regarded
as positive thanks to a sign symmetry in the theory, and through a
rescaling of the coordinates and of the other constant it can be
taken as
unit.\\
So, the event horizon is located at
\begin{equation}
    r_H=r_0+2l\sqrt{-\alpha}.
\end{equation}

\subsection{The n=4 solution}

In four space-times dimensions the solution is
\begin{align}
    A(r)&=e^2{(r-r_0)}^2+2\alpha;\\
    \phi(r)&=\frac{1}{r-r_0}.
\end{align}
The horizon manifold is a two-sphere, while there is a more
stringent requirement on the physical parameters than before:
$\Lambda=0$, $B^2=e^{-2}$, $\alpha<0$ in order to have an event
horizon. Again, the amplitude of the scalar field $c$ was scaled
away.\\
The horizon is located at radial coordinate
\begin{equation}
    r_H=r_0+\sqrt{-\frac{2\alpha}{e^2}}.
\end{equation}

\subsection{The n=5 solution}

The last example of this class of solution, and the one that could
actually represent he near-horizon approximation of a yet to be
found five dimensional black hole, is given by
\begin{align}
    A(r)&=\frac{2}{3}\Lambda{(r-r_0)}^2+\frac{8}{9}\alpha;\\
    \phi(r)&=\frac{1}{{(r-r_0)}^{3/2}}.
\end{align}
As before, the charge and cosmological constant are constrained,
$\Lambda=e^2$, while the direct product factor is given by
$B^2=3/2\Lambda$. The horizon is a three-sphere, located at
coordinate
\begin{equation}
    r_H=r_0+2\sqrt{\frac{\alpha}{3\Lambda}}.
\end{equation}
This expression forces the potential parameter $\alpha$ to be
positive and, as implicit also in the previous cases, constrains
$r_0$ to be not smaller than a certain value in order to have an
event horizon.

\section{Thermodynamics of black hole solutions}\label{thermodynamics}

In this Section, we shall study some thermal properties of the
hairy BHs, just as they follows from the form of the solutions. Recall
the Hawking temperature can be computed 
by means of \beq T_H=\frac{A'(r_H)}{4\pi}\,, \eeq and the  BH
entropy via Wald's method is \beq S=\frac{A_H}{4G}\at 1-\xi_c
\phi^2(r_H)\ct= 2\pi A_H\at 1-\xi_c \phi^2(r_H)\ct\,. \eeq Note
the presence of the  non standard and non  manifestly positive
factor depending on the scalar field. A naive application of the
method could lead to  problematic results because negative black
hole entropies might appear \cite{Cvetic:2001bk,Nojiri:2002qn,Cai:2003gr}.\\
However, for $\Lambda <0 $,  all the BH solutions found so far have  non
negative entropies. For $ \Lambda >0$, negative entropies may appear and
the $n=4$ case has been considered in \cite{barlow}.\\
Here we present a detailed investigation of the thermodynamics
associated with all BH hairy solutions
previously discussed. We are not going to consider the result for
non-warped solutions,
as they are not interesting to our purposes and there are no
ordinary solution to check if they are really a near horizon limit
of some general solution by means of a matching of the
thermodynamical parameters.

\subsection{The $n=3$ black hole solution with $\Lambda <0$}

In general, for $0< \omega <1$,  one has $r_H=f(\omega)r_0$ and
\beq
T_H=\frac{3 r_0}{2 \pi l^2}(1-\omega)\frac{1+f(\omega)}{f^2(\omega)}\,,
\nonumber
\eeq
where
\beq
f(\omega)=\at 1-\omega \ct^{1/3}\aq \at 1+\sqrt \omega \ct^{1/3}+
\at 1-\sqrt \omega \ct^{1/3}\cq \,.
\label{f}
\eeq
The entropy is  manifestly non negative
\beq
S_{BH}=2\pi A_H\frac{r_H}{r_H+r_0}= 4 \pi^2 \frac{f^2(\omega)}{1+f(\omega)}r_0\
\eeq
It is convenient to regard  $r_0$  as a  physical BH parameter.
Then,
taking the derivative of the BH entropy with respect to $r_0$
\beq
dS_{BH}=\frac{1}{T_H}dM \,,
\eeq
with the  identification
\beq
M=\frac{3\pi(1-\omega)}{l^2} r_0^2\,.
\eeq
One can check the validity of the Smarr relation
\[
M=\frac{1}{2}\,T_HS_{BH}
\]
It has been shown \cite{Banados:2005hm} that Smarr-like formulas can
be proved for many minimally coupled black holes in AdS, due to a
scaling symmetry of the reduced action; the remarkable universality of
these formulas, expressed by the independence of the scalar potential,
has been stressed by these authors. Here we see the validity of the
Smarr formula for conformal coupling as well. 
As a result, we get the first Law of the thermodynamics of the
black hole as well as the
 physical interpretation of the quantity $r_0$,
in  agreement with  \cite{Martinez:2004nb}, result obtained working in
the Einstein frame. Due to our theorem, both the Hawking
temperature and the BH entropy are  equal. We also have \beq
r_0=\frac{l}{\pi \sqrt{3(1-\omega)}}\sqrt{M}\eeq\,, \beq
T_H=\frac{\sqrt{3}}{2 \pi^2 l}(\sqrt{1-\omega})
\frac{1+f(\omega)}{f^2(\omega)}\sqrt{M}\,.
\eeq
\beq S_{BH}=\frac{4l}{\pi \sqrt{3(1-\omega)}}
 \frac{f^2(\omega)}{1+f(\omega)}\sqrt{M}=\frac{4l}{\pi \sqrt{3(1-\omega)}}
 \frac{f^4(\omega)}{(1+f(\omega))^2} T_H
\eeq
Thus, $M=g(\omega) T_H^2$ in agreement with the AdS/CFT correspondence.

Remark: Martinez et al.~\cite{Martinez:2004nb},  working in the
Einstein frame, 
got an expression for the Hawking temperature  formally different
from our. The two expression are identical (as they should) if and
only if 
\beq 
2\Im \left(\frac{\left( 1-\omega \right) ^{2/3}
\left( 1+i\sqrt {- \omega} \right) ^{2/3}}{\sqrt {-\omega}}\right)
=\frac{f^2(\omega)}{1+f(\omega)} \,. \eeq 
Only by a numerical
computation we were able to confirm it!\footnote{In the range
$\omega\in [-5,1]$ the numerical agreement was found to be of the
order of $4\cdot 10^{-9}$. No analytical method was found to compare
the two expressions.}

\subsection{The $n=4$ $\Lambda <0$ case}

In the $n=4$,  black hole solutions  exist only for $k=-1$, namely
for hyperbolic horizon. The Hawking temperature and the entropy
(which is positive) read \beq T_H=\frac{1}{2\pi
l}\sqrt{1+4r_0/l}\,, \eeq \beq S_{BH}=2\pi A_H\at
1-\frac{r_0^2}{(r_H+r_0)^2} \ct=2\pi V_Fl^2 \at
1+\frac{2r_0}{r_H}\ct=4\pi^2 V_F l^3 T_H \,, \eeq where $V_F$ is
the fundamental domain of the Riemann surface associate with the
hyperbolic horizon and BH entropy goes linearly with the Hawking
temperature.

If $r_0 >0$, there is  only an event horizon. If $r_0<0$, a finite
static region appears as well as an unbounded static region. This
is a quite unusual situation! The divergence of the scalar field
lies inside the inner static region, but the metric is otherwise
regular there.

As before, we may derive the First Law of the thermodynamics of
the black hole and determine the mass as a function of the
parameter $r_0$, taking the derivative with respect to $r_0$, \beq
dS_{BH}=\frac{1}{T_H}d\at 2V_F r_0\ct \,, \eeq \beq M=2r_0\,V_F
+M_0 \,, \eeq where $M_0$ is a constant independent on $r_0$. We
may determine  the constant, requiring $T_H$ to be real and the
mass non negative. Thus, \beq
M_0=-\frac{l\,V_F}{2}\,, \eeq and \beq M=2\pi^2V_Fl^3 T^2_H \eeq
Again one has the Smarr formula $M=T_HS_{BH}/2$.
Remark:  The AdS/CFT correspondence would suggest, at least for
large mass, \beq M=B T_H^3 \,. \eeq As a result, it seems that we have
an example of
violation of AdS/CFT correspondence. Furthermore, the entropy as a
function of the mass reads \beq S_{BH}=\pi(2 l)^{3/2}\sqrt{MV_F}
 \,, \eeq again in disagreement with the usual AdS/CFT
correspondence. However there exists a generalized AdS/CFT
correspondence \cite{Minces:2001zy} which can be adapted to this
case, but it remains to show how the explanation of the apparent
disagreement really works.  
The Hawking temperature and the entropy are equal
to the ones computed in the Einstein frame. The specific heat of
these solutions  is positive, namely there is thermodynamic
stability of the BH.

\subsection{The $n=4$ $\Lambda > 0$ case}

This is a  quite interesting case. It has been discussed for a charged BH
in \cite{barlow}. We anticipate that our conclusions will be slightly
different.  For the sake of simplicity, we consider only the uncharged
case, since our results can be easily extended to the charged one.

In order to deal with  event
horizon $r=r_E$ and a cosmological horizon $r_C$, the integration
constant $r_0$  must be negative, say $r_0=-a, \, a>0$. It turns
out that  event and cosmological horizon  have the same Hawking
temperature, lukewarm BHs  \beq T_E=T_C=\frac{1}{2\pi
l}\sqrt{1-4a/l} \,. \eeq Note that it  might be possible another
interpretation} for $ T_C$, due to Klemm and Vanzo \cite{Klemm:2004mb},
in term of  negative temperature. We will not pursue this point of
view.

In \cite{barlow} is  reported  that the cosmological BH entropy
is  positive and the other one associate with the event horizon is
negative, total entropy being  vanishing, \beq S_C=4\pi^2 V_F l^3
T_C \,, \eeq \beq S_E=-4\pi^2 V_F l^3 T_E \,. \eeq One can make
use of  the First Law of BH thermodynamic and get \beq
a=\frac{M}{8\pi} \,, \eeq where $M$ may be interpreted as  BH
mass. The temperatures may be rewritten as \beq
T_E=T_C=\frac{1}{2\pi l}\sqrt{1-\frac{M}{2\pi l}} \,. \eeq As a
result, a maximal mass and temperature are present: $M < 2\pi l$.

A negative entropy, in the statistical sense, does not
make sense, so either one removes by hand the negative part or else one
interprets the negative entropy as meaning that the lukewarm states
have exponentially small probability, because in de Sitter space there
are robust arguments to believe that the entropy is the logarithm of
the euclidean action (see \cite{Iyer:1995kg} for a comparison of
Noether charge with euclidean methods). However, in de Sitter space it
could be that there is not a universal state with zero entropy to
which compare all other states, and there is some evidence of this in
$3D$ de Sitter space \cite{KV}. In this case only entropy differences
make any sense. 
As emphasized by 't Hooft \cite{'t Hooft:1984re}, it could also
be the case that the additive constant diverges if the contribution
of the quantum fields around a black hole are not carefully
renormalized. So shifting the entropy by a constant not 
only is a sensible procedure in quantum gravity, it is also a
transformation that can be given a precise meaning within the Noether
charge method, as discussed extensively in \cite{Clunan:2004tb}.

This motivates our proposal: make use of the ambiguity in the Wald
method and fix
the constant in the entropy by the following continuity argument.
When $M \rightarrow 0$, one  must get the pure de Sitter solution.
Thus, a reasonable expression for the entropy associated with the
event horizon is \beq\label{ennew} S_E=2\pi V_F l^2 \at
1-\sqrt{1-\frac{M}{2\pi l}} \ct \,. \eeq With this proposal, the total
entropy is not vanishing and equal to the de Sitter one \beq
 S_{dS}=2 \pi V_F l^2=8 \pi^2 l^2 \label{}
\eeq
Our proposal can be strengthened by looking at the Smarr
formula. Using the ``renormalized'' entropy \eqref{ennew}, with a 
direct verification one has 
\beq\label{smarr}
\frac{1}{2}\,(T_ES_E+T_CS_C)=\frac{1}{4\pi l}\at\sqrt{1-\frac{M}{2\pi
    l}}\ct 
\eeq
and it is easily seen that the last term is the vacuum energy in between
the two horizons, namely $(4\pi)^{-1}\int\Lambda K_ad\Sigma^a$, where
$K$ is the timelike Killing field of the solution, in agreement with
\cite{hg}. But looking to \eqref{ennew}, we see that this entropy does
not follows anymore the area law.

\subsection{The n-dimensional de Sitter case}

We conclude this Section with the $n$ dimensional de Sitter
solution. For a generic $n$ and $k=0$, we have shown that there is
a de Sitter solution of our gravity  conformally coupled to a
scalar hair. In the static gauge, there is a cosmological horizon
and the scalar field is non static. However, the time dependence
disappears on the horizon. Thus, if we compute the entropy with
the Noether charge method, we have \beq S_{BH}=2\pi V_F l^{n-2}\at
1-\xi_c\frac{1}{r_0^{n-2}} \ct  \,, \eeq or \beq S_{BH}=2\pi V_F
\aq l^{n-2}- \xi_c\at \frac{n-2}{2\sqrt{2|\alpha_n|}}\ct^{n-2}
\cq\, . \eeq The first contribution is the pure de Sitter
contribution, the second one is constant  and  negative and
associated with the degrees of freedom of the scalar field.

As a result, as it is well known,  pure de Sitter space
is the state of  maximum entropy.

\section{Conclusions}

In this paper, we have revisited the Einstein gravity-scalar
system in various space-time dimensions, with a  conformally
coupled scalar field and non vanishing cosmological constant.
Motivated by RS models, we have worked in the so called Jordan
frame. First, we have provide an elementary proof of the conformal
frame independence of Hawking temperature and black hole entropy.
Motivated by this result, we have revisited the search for exact
n-dimensional black hole solutions in presence of the conformally
coupled scalar field. We have recovered the $n=3$ and $n=4$ known
solutions and found a  new de Sitter n-dimensional solution with
time dependent scalar field. The thermodynamics properties of
these hairy black hole solutions have been discussed in detail and
their entropy evaluated with the Noether charge method, which is
powerful but does not ensure the positivity of the entropy.

For asymptotically AdS hairy BH solutions, the related BH
entropies turn out to be non negative and  this fact is certainly 
related to their  stability.

For asymptotically dS hairy BH solutions, a naive evaluation of the BH
entropy leads to a negative entropy. Making use of the freedom in the
Wald method, a simple continuity argument has been advocated in order
to remedy this fact. The resulting expression amusingly satisfy the
Smarr relation pertinent to de Sitter space.

As far as the AdS/CFT correspondence, in the negative cosmological
constant case, we have seen that the $n=3$ BH solution is
in agreement  with it, while  the $n=4$ BH solution, gives rise to a
relation between the mass and Hawking temperature in disagreement
with the usual  AdS/CFT correspondence result. The Minces/Rivelles
generalized AdS/CFT correspondence is probably the way out of this
dilemma.

\end{document}